\documentclass[aps,preprint,groupedaddress,showpacs]{revtex4}
\usepackage{graphics}
\usepackage{graphicx}
\usepackage{amsmath}
\usepackage{amssymb}
\usepackage{bm}
\usepackage{dcolumn}
\begin{document}
  \preprint{version 1.0}
 \title[]{Metastable liquid lamellar structures in 
binary and ternary  mixtures of Lennard-Jones fluids}

\author{Enrique D\'\i az-Herrera$^{a}$,
Guillermo Ram\'\i rez-Santiago$^{b}$
and Jos\'e A. Moreno-Razo$^{a}$}
    \affiliation{$^{a}$Departamento de F\'\i sica,
     Universidad Aut\'onoma Metropolitana-Iztapalapa,
     Apartado Postal 55-534, M\'exico 09340, D.F., MEXICO\\
$^{b}$Instituto de F\'{\i}sica, Universidad Nacional 
           Aut\'onoma de M\'exico, Apartado Postal 20-364, 
           M\'exico 01000, D. F.,  MEXICO}

\begin{abstract}
We have carried out extensive equilibrium molecular dynamics (MD)
simulations to investigate the Liquid-Vapor  
coexistence in partially miscible binary and ternary mixtures
of Lennard-Jones (LJ) fluids. We have studied in detail the time evolution of
the density profiles and the interfacial properties in a temperature 
region of the phase diagram where the condensed phase is demixed. 
The composition of the mixtures are fixed, 50\% for the binary mixture 
and 33.33\% for the ternary mixture. The results of the simulations 
clearly indicate that in the range of temperatures $78 < T < 102~^{\rm o}$K,
--in the scale of argon-- the system  evolves towards a metastable 
alternated liquid-liquid lamellar state in coexistence with its vapor phase. 
These states can be achieved if the initial configuration is fully disordered, 
that is, when the particles of the fluids are randomly
placed on the sites of an FCC crystal or the system is completely mixed. 
As temperature decreases these states become very well defined and more 
stables in time. We find that  below $90~^{\rm o}$K, the alternated 
liquid-liquid lamellar state remains alive for 80 ns, in the scale of argon, 
the longest simulation we have carried out. 
Nonetheless, we believe that in this temperature region these states will 
be alive for even much longer times. 
\end{abstract}
\pacs{PACS numbers: 61.20.Ne  61.20.Ja  61.25.Em  61.20.Gy }

\maketitle


\section{introduction}
\label{introduction}
In the past decade there has been a growing interest in studying the 
structure and interfacial properties of the  liquid-liquid (LL) and 
liquid-vapor (LV) phase coexistence of binary and ternary mixtures of 
simple fluids. 
The reason for this is that they are of fundamental importance 
to understand the behavior of more complex fluids.
A few  examples where a better understanding of these phenomena
may have impact are, in the design, operation and optimization of the 
extraction processes, in tertiary oil recovery, in the adsorption or 
coating processes, in biological processes, etc. 
Different analytical approaches
\cite{dagamma1,tarazona1,jlee1,almeida}, and density functional
theory implementations \cite{wendall,fortsman97,nader2002},
as well as  numerical 
simulations\cite{chapela,salomons,holcomb,li95,mecke99,alejandre99,diaz99} 
have been used  to investigate these issues in simple fluids and in more 
complex systems~\cite{strey95,guerra98,endo2000,isamu2001}.
Binary mixtures of LJ
fluids have been investigated with the aim at understanding the phase 
coexistence as well as the interfacial and structural properties 
of the LV and LL interfaces. 
The dependence of the surface tension on the composition
of the bulk phases and surface segregation at the LV 
interface have also been analyzed~\cite{jlee1,salomons,nijmeijer88}.
For two immiscible LJ fluids the structure of the LL interface, the 
capillary waves and  pressure effects on the interfacial tension
have also been studied~\cite{toxvaerd95,stecki95,padilla96}.
In addition, the structure of LV and LL interfaces in binary mixtures 
of LJ fluids have been investigated using density functional 
theory~\cite{fortsman97} as well as molecular dynamics 
simulations~\cite{mecke99}.
In the present paper we have investigated the time evolution of the density
profiles and interfacial properties of  binary and  ternary mixtures of
partially miscible LJ fluids in a temperature region of the phase diagram
where the condensed phase is demixed.
The phase diagram that better describes our model binary mixture 
is one that has the usual low temperature triple point and a higher
temperature triple point where the vapor
phase coexists with a mixed and a demixed liquid  phases~\cite{wilding97}.
These types of mixtures have received attention
until recently and need to be studied more extensively. Our findings
indicate that by properly tuning the temperature, 
$78 < T < 102~^{\rm o}$K, --in the scale of argon-- and the
interactions between the species, 
long lived metastable alternated lamellar states in the liquid-liquid phase
coexist  with the vapor phase. These states are found in binary and ternary 
mixtures at compositions of  50\%, and 33.33\%, 
respectively. As temperature decreases these states
become sharper and more stables in time. We find that  below
$90~^{\rm o}$K, the alternated liquid lamellar state remains alive for at 
least 80 ns in the scale of argon. This is the longest simulation that
we have carried out. However, we believe that below this temperature these
states will remain there for even very long times, suggesting that 
{\it the alternated liquid-liquid phases are a manifestation of the 
existence of 
a number of local equilibria.} The layout of the 
remainder of this paper is as follows. In section \ref{models} we describe
the model potentials that define the type of mixtures that we have investigated.
Then we continue in  section \ref{sim-deltails} with  a description
of the parameters of the MD simulations. The results and discussion of them 
are presented in section \ref{results}. Finally, we end the
paper with the conclusions that are in section \ref{conclusions}.
\section{The models.}
\label{models}
Let us start this section by defining the model system for the binary fluid 
mixture. This is made up of spherical particles A and B that 
interact between themselves through a LJ potential defined by
\begin{equation}
u_{ij}=4\epsilon_{ij}\left[ \left( \frac{\sigma_{ij}}{r_{ij}}\right)^{12}
- \left( \frac{\sigma_{ij}}{r_{ij}}\right)^{6}\right], 
\label{pot}
\end{equation}
where i,j=A,B. We will use  $\sigma_{AA}$ and $\epsilon_{AA}$ as the 
reference parameters, that is,  $\epsilon_{ij}=\epsilon_{AA}\alpha_{ij}$ and 
$\sigma_{ij}=\sigma_{AA}=1.0$. 
This choice of parameters represents a system in which all the 
particles are of equal size and the interactions are
tuned by properly choosing the elements of the symmetrical matrix 
$\alpha_{ij}$. Since we want to study a partially miscible
binary mixture the attractive part of the interaction potential 
between particles of fluids A and B is chosen to be weaker than the
attraction between particles of the same specie. 
Therefore all the matrix elements $\alpha_{ij}$ are equal 
to unity except those corresponding to the crossed AB 
interactions that are $\alpha_{AB}$=0.75. This choice of potential 
favors the demixing of fluids A and B.
On the other hand, the ternary mixture we have considered is 
formed by fluids A, B and C which particles are of the same size 
for the three fluid components.
The potential interactions between fluids of different kind are chosen such 
that they favor the mixing of the pairs of fluids B,C and A,C (alike 
interactions). However, the A-B fluids interactions disfavor 
their mixing (unlike interactions). Bearing this in mind we can define
the interaction potential in a similar way as we did for the binary
mixture. It is defined by Eq. \ref{pot} where
i,j=A,B,C, and the matrix elements $\alpha_{ij}$ are equal to unity for
particles of the same type as well as the crossed interactions
BC and AC. Nonetheless, the AB crossed interactions have
matrix elements $\alpha_{AB}=0.75$. 
With this choice of potentials, fluid C may be considered  as a surfactant.
The reason for this is that particles of fluid C mediate the unlike A-B 
fluids interactions, that is, particles of fluid C  interact favorably
with particles of fluids A and B, while particles of 
fluids A and B interact unfavorably~\cite{diaz99}.
In the   following section we present information concerning 
the parameters and details of the numerical simulations.
\section{Parameters and description of the simulations.}
\label{sim-deltails}
We have carried out extensive equilibrium molecular dynamics simulations
in the (N,V,T) ensemble.  In each time step iteration we monitor the 
temperature of the system via the equipartition theorem and rescale 
the linear momentum of the particles to keep the temperature constant.
In all the simulations performed in this work 
the interparticle potentials have been shifted in such a way that they are 
equal to zero right at the cutoff $r_{c}=3\sigma_{AA}$. It has been shown 
that this cutoff is the appropriate to account for the long tail of the 
potential~\cite{alejandre99}.
We have simulated systems with  as many as 4096 particles for the binary
mixture and  8192 particles for the ternary mixture.
They  are located inside  a parallelepiped  of volume
$12\sigma_{AA}\times 12\sigma_{AA}\times L_{z}$, with $L_{z} > 12\sigma_{AA}$.
We have applied periodic boundary conditions in all three directions. 
To minimize finite size effects in the interfacial properties we have
followed reference~\cite{li95}. There it is shown that in order to have
interfacial properties independent of the size of the area it should be 
at least $12\sigma_{AA}\times 12\sigma_{AA} $.
We have found that for the type of mixtures --interactions-- studied in
this work the choice of the initial configuration is crucial so we leave 
this discussion for the next section.  At this point
all we can say is that the initial velocities of the particles are  chosen 
from an equilibrium Boltzmann distribution.
The thermodynamic variables used to describe the system's behavior are, the
reduced  temperature $T^{*}= \frac{k_{_B} T}{\epsilon_{AA} }$ with $k_{_B} =$
Boltzmann's constant and the reduced density 
$\rho^*={ {\rho}{\sigma_{AA}^{3}} }$. To investigate the interfacial 
properties we need to define the reduced interfacial tension,
$ \gamma^{*} = (\sigma^{2}_{AA}/\epsilon_{AA})\gamma$.
The total density of the system is given by $\rho=N/V$, where $N$ is
the total number of particles and $V$ is the volume of the system.
The equations of motion that describe the dynamics of the particles of the 
fluids are solved using a leap-frog scheme with an integration 
step-size $\Delta t^{*}\le 5\times 10^{-3}$. 
To be able to compare the time scale and temperatures of our 
simulations in the remaining of this paper we will use as a 
reference scale that corresponding to argon. In this scale the 
time step is equal to  $1.1\times 10^{-5}$ nanoseconds (ns). 
The shortest equilibration times  that we considered in the simulations were
of the order of $5\times 10^{5}$ time steps (5.5 to 11 ns)
and the total length of the simulations after equilibration is in the range 
of 4-7 million time steps (5.5 to 11 ns).
To minimize correlations between measurements we
calculated thermodynamic and structural quantities every 50 time steps.
Since we want to study the liquid-vapor phase coexistence the
total density of the system has been chosen to be right inside  the LV 
coexistence curve. In this way we are able to study directly the structural
properties of the interfaces and explicitly obtain the distribution of the 
species in the directions parallel and perpendicular to the interfaces.
\section{Results and discussion.}
\label{results}
\begin{figure}
\begin{center}
\includegraphics[width=3.25in,height=1.25in]
{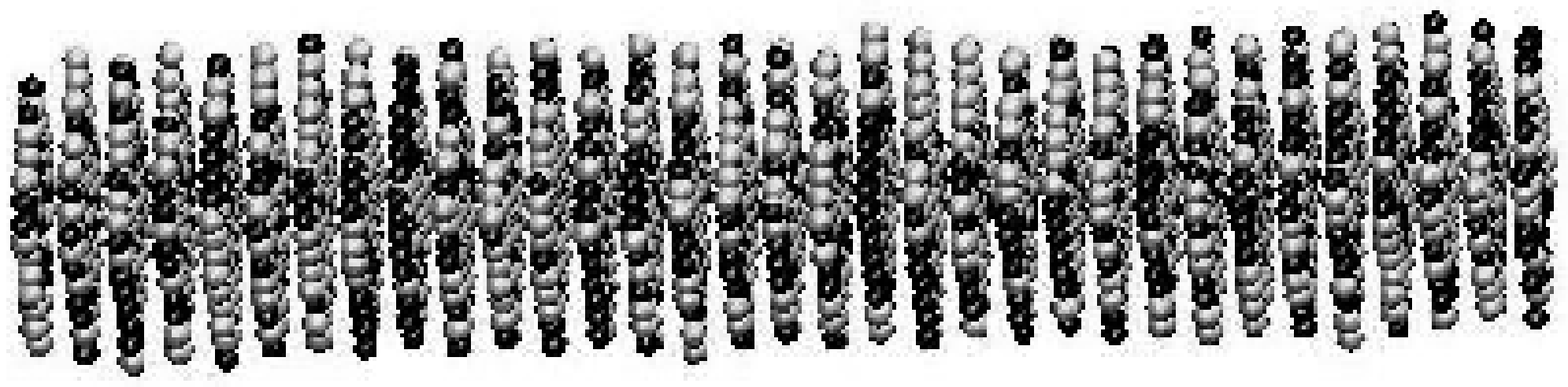}
\end{center}
\caption{Snapshot of the binary mixture {\it ``disordered initial
configuration"}.  At $t=0$ the  particles of fluids A (black) and
B (gray) are randomly located on the sites of an FCC crystal.}
\label{fcc}
\end{figure}

\begin{figure}
\begin{center}
\includegraphics[width=3.25in,height=1.25in]
{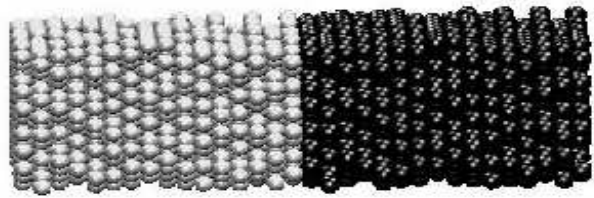}
\end{center}
\caption{Snapshot of the binary mixture {\it ``ordered initial
configuration"}. At $t=0$ the  particles of fluids
A (black) and B (gray) are separated and located on the sites of two
adjacent FCC crystals.}
\label{sep-fcc}
\end{figure}

We would like to recall that our main objective is to focus on the 
time evolution of the demixed liquid-liquid-vapor (LLV) phase coexistence 
of the binary mixture whose phase diagram is described in
reference~\cite{wilding98}. These types of mixtures have received much less
attention than those based on the Lorentz-Berthelot rule that usually
yield a mixed liquid phase. We have found that in the partially miscible
mixtures studied here, care should be exercised when choosing the initial
configuration of the simulations since there are long lived metastable states. 
It is known that an equilibrium state can be reached from any two different 
initial configurations. Nonetheless, for practical purposes, one always 
chooses an initial configuration that is close to the expected equilibrium 
state. So, in the case of a partially miscible mixture it is natural to start 
from a configuration in which the species are separated if one expects 
to obtain a demixed liquid state. On the contrary, if in the final equilibrium 
state the liquids are mixed then one should initiate the simulation from a 
mixed configuration. In this paper we show evidence that long lived 
metastable alternated liquid-liquid lamellar states  in coexistence with the
vapor phase can be achieved if the simulations start from a {\it ``disordered
initial configuration"} defined as one in which fluids A and B
are fully mixed in the liquid-liquid phase or their particles are randomly
located on the sites of an FCC crystal, as it is shown in Fig.~\ref{fcc}.
However, if we start  the simulations from an
{\it ``ordered initial configuration"} defined as one in which the particles
of fluids A and B are placed on the sites of two adjacent FCC crystals
or they are distributed on two contiguous slabs containing fluids A and B 
respectively,  as shown in Fig.~\ref{sep-fcc}, the system ends up in 
an equilibrium configuration where the distribution of fluids essentially 
remains the same as in the initial configuration.

To allow the formation of the vapor phase in the
MD simulations we have left a free volume on both sides
of the initial distribution of particles. However, this empty space is not
shown in the snapshots. 
In the next two subsections we show and discuss the results of the time
evolution of a binary and a ternary mixtures.
\subsection{Binary mixture.}
\label{binary}
\begin{figure}
\begin{center}
\includegraphics[width=3.25in,height=3.0in]{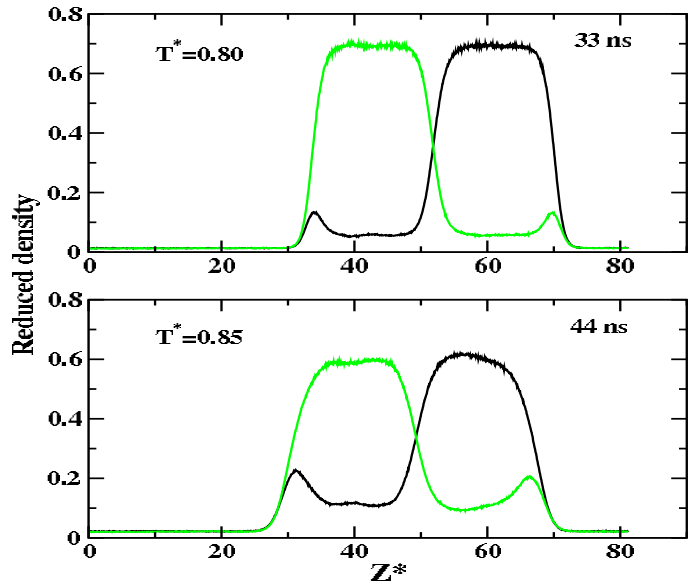}
\end{center}
\caption{Density profiles of the binary mixture in the demixed liquid
state in coexistence with the vapor phase. Notice that the density profiles
of the fluid phase that is less rich  show a peak structure close to the
LV interface. See text for an explanation of this result.}
\label{2liq}
\end{figure}

\begin{figure}
\begin{center}
\includegraphics[width=3.25in,height=3.0in]{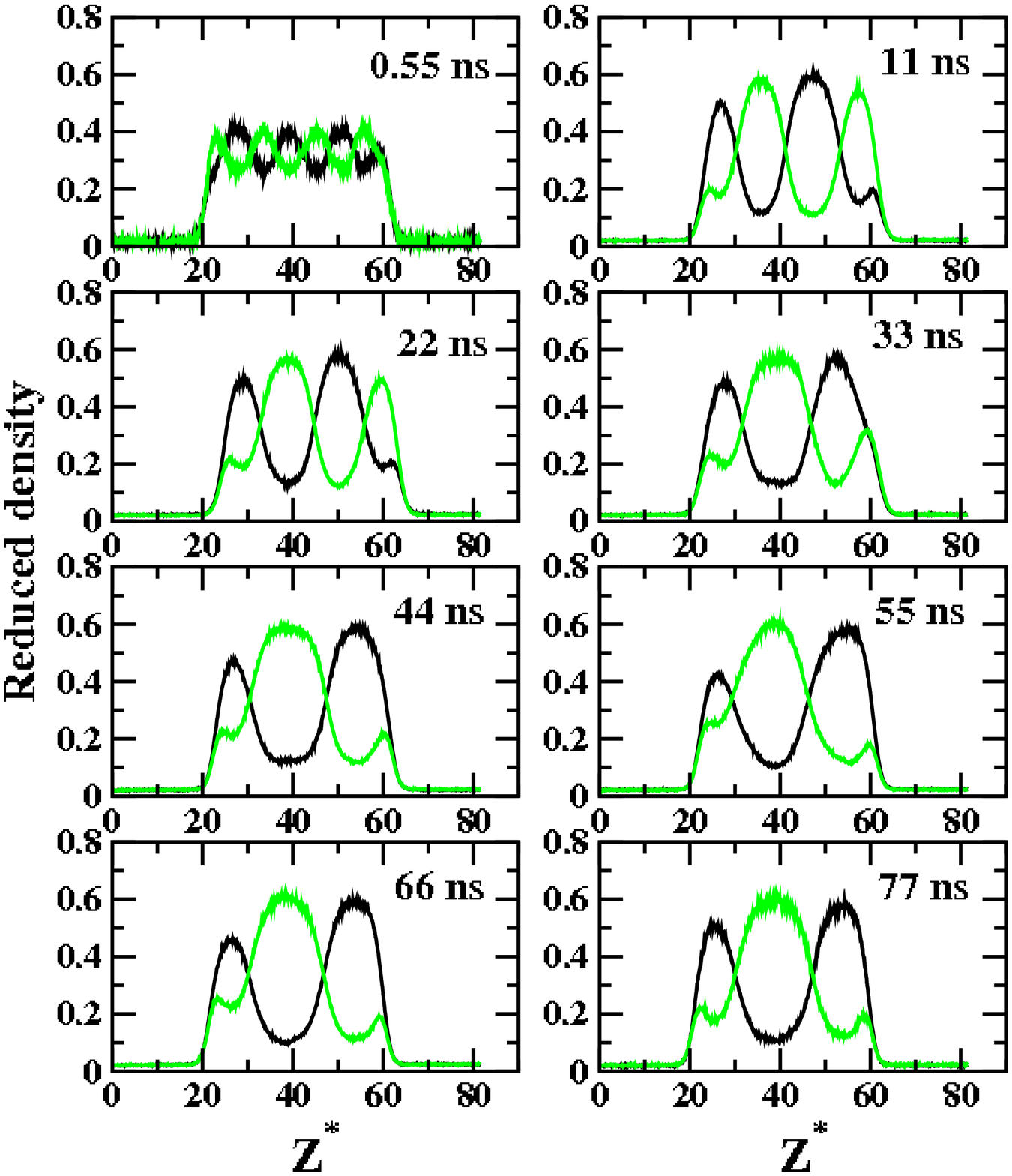}
\end{center}
\caption{Density profiles for the mixed condensed state of the system
in coexistence with its vapor phase at $T^{*}=0.85$.}
\label{lam85}
\end{figure}

\begin{figure}
\begin{center}
\includegraphics[width=3.25in,height=3.0in]{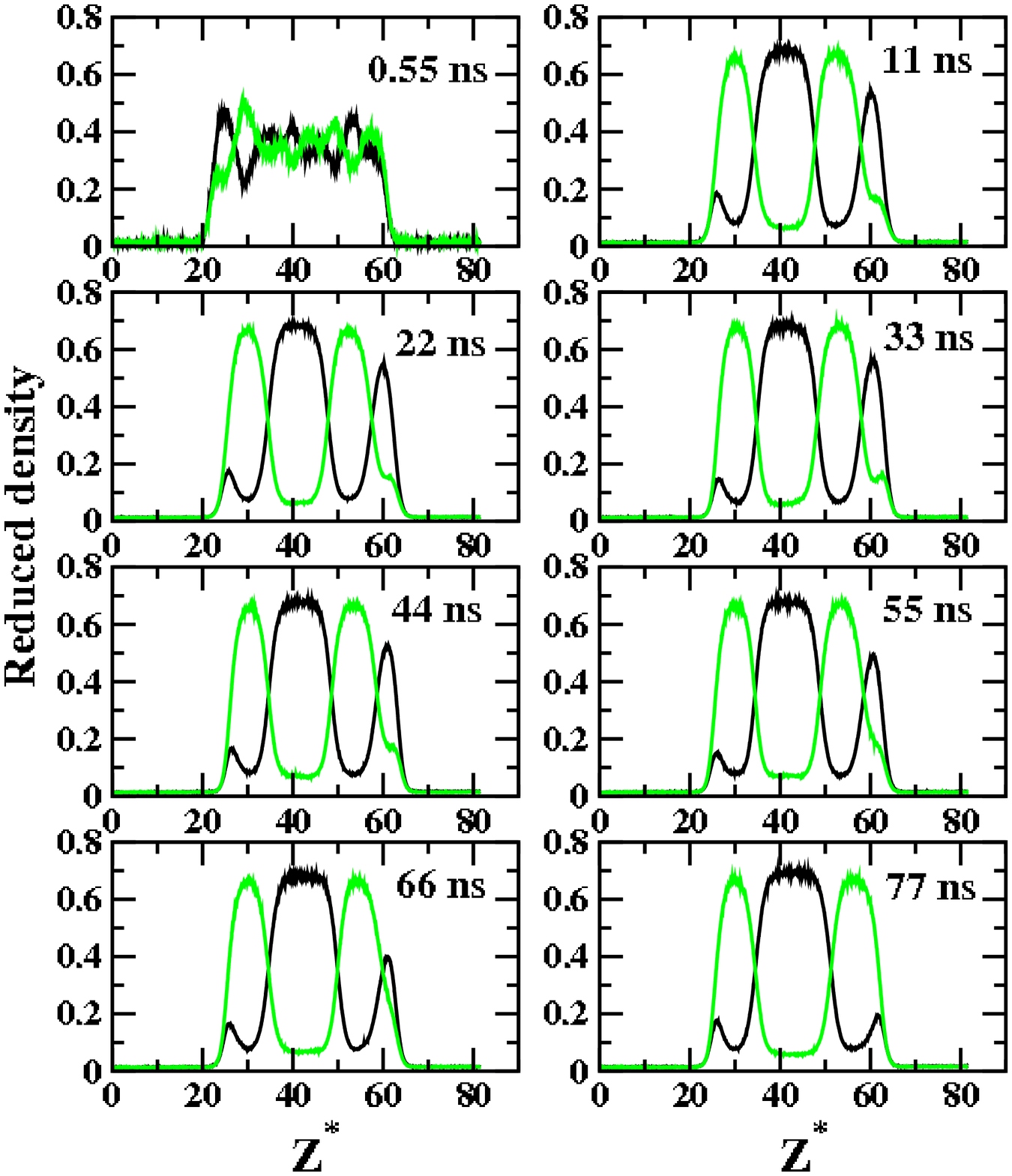}
\end{center}
\caption{Density profiles for the mixed condensed state of the system
in coexistence with its vapor phase at $T^{*}=0.80$.}
\label{lam80}
\end{figure}

Let us start by considering a typical LLV interface structure when the 
simulation begins from an ``ordered configuration" as that shown in
Fig.~\ref{sep-fcc}. The mixture has a total of 4096 particles with a
concentration of fluids A and B of 50\%.  We have considered an average 
particle density of $\rho^{*}_{\rm tot.}=\rho^{*}_{A}+ \rho^{*}_{B}= 0.35$.
In figure~\ref{2liq} we show the density profiles of the LLV coexistence at
the reduced temperatures $T^{*}=0.85$ and $0.80$. In our reference scale
these temperatures correspond to $T=101.83~^{\rm o}$K, and 
$T=95.84~^{\rm o}$K, respectively, and are in the demixing region of 
the phase diagram~\cite{wilding98}. Therefore we obtain two partially 
separated liquid phases, one  rich in fluid
A and the other rich in fluid B and both of them in coexistence with the vapor 
phase. Notice that this is the minimum number of interfaces that one can 
obtain using periodic boundary conditions.  
Observe that in the phase of fluid A or B  one can also find fewer
amounts of fluid B or A, whose average density is small and constant in the
bulk phase and develops a peak structure at the LV interfaces as can be seen
in density profiles. This distribution of particles  occurs as a consequence
of the difference in the local densities and the unlike potential
interactions between the fluids. It is important to point out that this
kind of structure can only be obtained by means of the construction of the
interfaces as we have done in the present MD simulations.
Another important fact to bear in mind is the amount of simulational time
required to obtain the equilibrium density profiles such that the two
liquid phases are completely symmetric. So we needed to simulate
these systems for as long as  33~ns and 44~ns at each one of these two 
temperatures, respectively. 
On the other hand, figures \ref{lam85} and \ref{lam80} show the time evolution 
of the density profiles of the mixture at the same temperatures, 
$T^{*}=0.85$ and $0.80$. When the simulations started from a
``disordered initial configuration" as the one shown in Fig.~\ref{fcc} we
observe that at both temperatures, and at the early stages of the time 
evolution, $11$~ns, the system clearly shows the signature of an alternated
A-rich and B-rich liquid lamellar structure in coexistence with the vapor 
phase. One would expect that the number of lamellae should be proportional 
to the number of particles in the system.
To test the time stability of these alternated structures we
performed quite long MD simulations, as long as $77$~ns for each temperature. 
The time length of these simulations is about 1.75 to 2.3 times longer than 
the time we followed the evolution of the systems initiated from an ``ordered 
initial configuration", that in turn yielded the structure shown in 
Fig.~\ref{2liq}. We find that at high temperatures the lifetime of the
lamellar structure is relatively short because the kinetic energy overcomes
easily the LL interfacial energy barrier. In Fig.~\ref{lam85} 
it is seen that for $T^{*}=0.85$ one of the lamellae has dissapeared at 33~ns.
Nonetheless, at  a lower temperature $T^{*}=0.80$ the time required for this 
disappearance to occur is between 66-77~ns, about two times longer than
at $T^{*}=0.85$. This  is  due to the diffusion of particles from the 
narrowest lamella that becomes unstable, in this case, the one on the 
right hand side, towards the lamella at the middle, that in turn becomes 
thicker. The fact that the right hand side lamella is the narrowest is due to 
a local density fluctuation in the system.  However, this local density 
fluctuation may also occur in any one of the lamellae.   
To understand why this diffusion of particles occurs in this direction 
and not in the direction of the vapor phase we estimated the LV and LL 
interfacial energies at both temperatures. In doing so we integrated the 
difference between the normal and tangential pressure profiles~\cite{diaz99}.
One more test of the reliability of the simulations is to check that this
difference is zero in the bulk phases and that the normal pressure is
constant through the length of the simulational parallelepiped.
Our results indicate that the LV interfacial tensions  
$\gamma^{*}_{LV}(T^{*}=0.85)=0.290\pm 0.005$
and $\gamma^{*}_{LV}(T^{*}=0.80)=0.420\pm 0.006$) are significantly
greater than those corresponding to the LL interfacial tensions
($\gamma^{*}_{LL}(T^{*}=0.85)=0.07\pm 0.01$ and 
$\gamma^{*}_{LL}(T^{*}=0.80)=0.150\pm 0.001$). Consequently, 
the fluid particles  forming the lamella at the right end of the LL structure
have to diffuse towards the liquid lamellar phase since more energy is 
required to move towards the vapor phase. 

\begin{figure}
\begin{center}
\includegraphics[width=3.25in,height=3.0in]{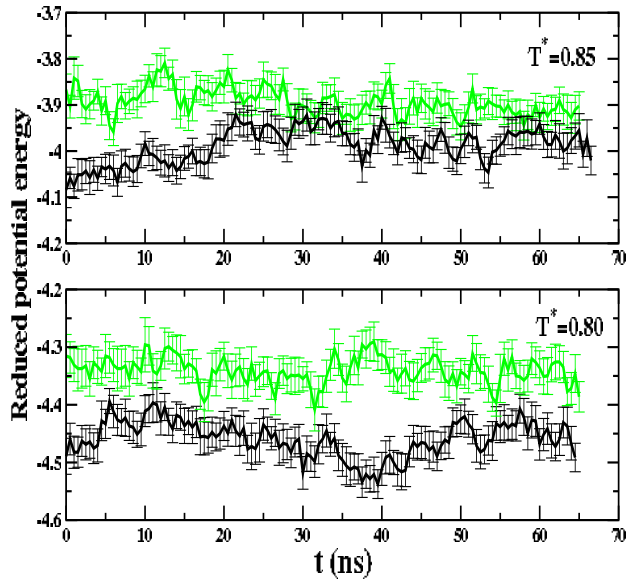}
\end{center}
\caption{Time evolution of the potential energy of the mixture
at reduced temperatures $T^{*}=0.85 \>{\rm and}\> 0.80$. The data in
gray color correspond to a demixed fluid phase configuration while
the data in black correspond to an ordered initial configuration.}
\label{upot2spec}
\end{figure}

\begin{figure}
\begin{center}
\includegraphics[width=3.25in,height=3.0in]{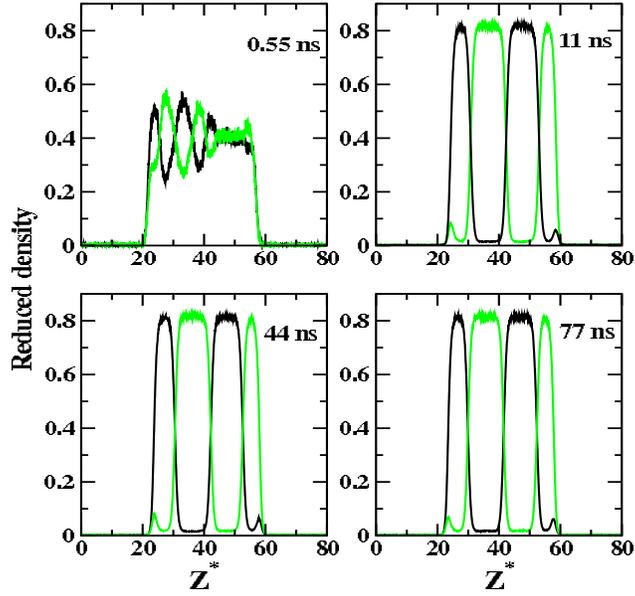}
\end{center}
\caption{Density profiles for the mixed condensed state of the system
in coexistence with its vapor phase at $T^{*}=0.65$. Note that in the bulk phase, but
close to the LV interface, the density profile of fluid C shows
peaks as a consequence of the unlike interactions between the A  and
B particles.}
\label{lam65}
\end{figure}

\begin{figure}
\begin{center}
\includegraphics[width=3.25in,height=1.25in]{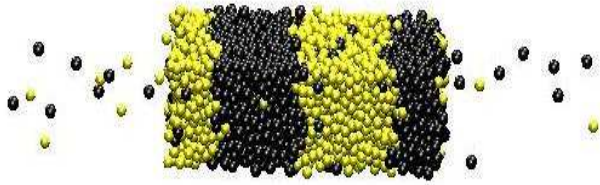}
\end{center}
\caption{Snapshot of the particles arrangement in the long lived
alternated lamellar state in coexistence with its vapor phase at
$T^{*}=0.65$ at 77~ns of the simulation.}
\label{snapshot65}
\end{figure}

To show that the system at $T^{*}=0.85$ has a shorter lifetime and eventually 
will end up forming two liquid slabs in coexistence with the vapor phase  we 
followed the time evolution of the potential energies when the simulations are  
started from an ordered and a disordered configurations. 
The results are plotted in Fig. \ref{upot2spec}
where one sees that for $T^{*}=0.85$ the behavior of the potential energy
is such that after about $t_{a}=27.5$ ns both curves are very close to each 
other within the size of the error bars. This indicates that for $t>t_{a}$ 
the potential energies of  both systems are the same within the accuracy of 
the simulations, suggesting that in both cases the system is approaching  the
same thermodynamic state. However, this appears not to be the case for 
$T^{*}=0.80$ in which case
the time evolution of the potential energy indicates that the curves are
clearly separated when the initial configurations are different. Thus one
can say that the system may converge to two possible states. One in which
the free energy is minimum, --just one LL interface-- and the other with
many LL interfaces --alternated lamellar state-- whose free energy appears not
to be the minimum due to the existence of several interfaces.
As we decrease further the temperature
we find that the liquid lamellar state becomes very well defined --sharper-- 
and more stable. In Figure~\ref{lam65} we show the time evolution of the 
density profiles at the reduced temperature $T^{*}=0.65$ where once more, 
the simulations started  from a  disordered configuration.
It is important to note that if the simulation starts
from an ordered configuration then the system ends up forming two
separated liquid slabs in coexistence with the vapor phase.
At this temperature the alternated lamellar structure is very well defined
at early times of the simulations ($t=11$~ns). Notice that in this case 
the LL interfaces are sharper and the width of the lamellae is narrower and
very well defined as compared to those obtained at higher temperatures.
Note that the simulations are quite long, $77$~ns and the structure is stable
during all this time. This is not totally unexpected since as temperature
decreases the diffusion of the particles of fluid A in the phase of fluid B 
decreases as a consequence of the values of the interfacial tension of the
LL interface, $(\gamma^{*}_{LL}=0.46\pm 0.02)$. 
This value is about 6.6 times greater than the value of $\gamma^{*}$ at 
$T^{*}=0.85$. In Fig.~\ref{snapshot65} we show a snapshot of the system at 
$T^{*}=0.65$ after 77 ns of the simulation. There one can see explicitly the
lamellar structure whose density profiles are shown in the fourth panel of 
Fig.~\ref{lam65}.
In the next subsection we address the issue of how a surfactant fluid
--third specie-- affects the stability of the
lamellar state of the binary mixture at $T^{*}=0.65$.
\subsection{Ternary mixture}
\label{ternary}
\begin{figure}
\begin{center}
\includegraphics[width=3.25in,height=3.0in]
{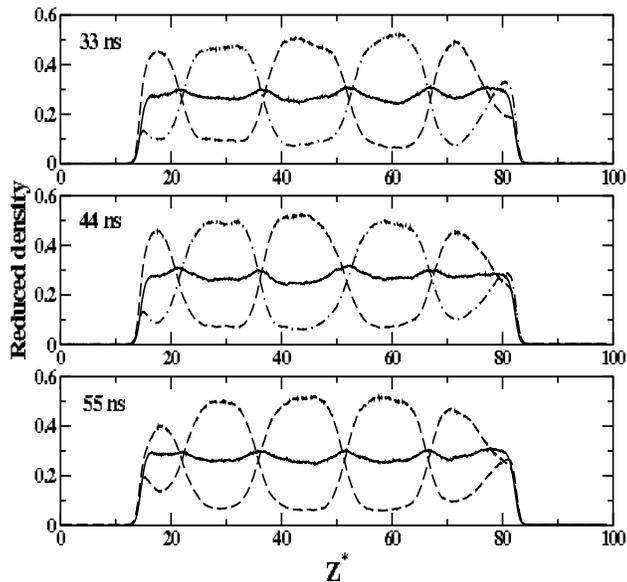}
\end{center}
\caption{Time evolution of the density profiles of the lamellar state at
$T^{*}= 0.65$. Averages are calculated using three subsequent blocks of
one million time steps each. The initial configuration is disordered.}
\label{laminas3spec}
\end{figure}

\begin{figure}
\begin{center}
\includegraphics[width=3.25in,height=3.0in]{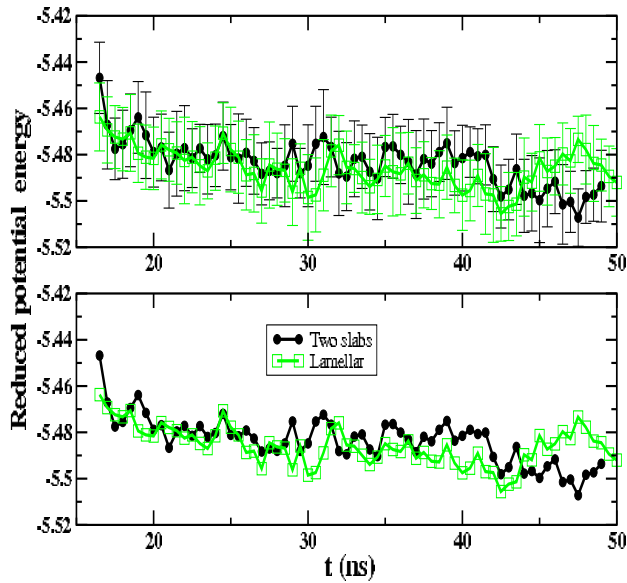}
\end{center}
\caption{Time evolution of the potential energy of the mixture
at $T^{*}=0.65$.}
\label{upot3spec}
\end{figure}

\begin{figure}
\begin{center}
\includegraphics[width=3.25in,height=3.0in]
{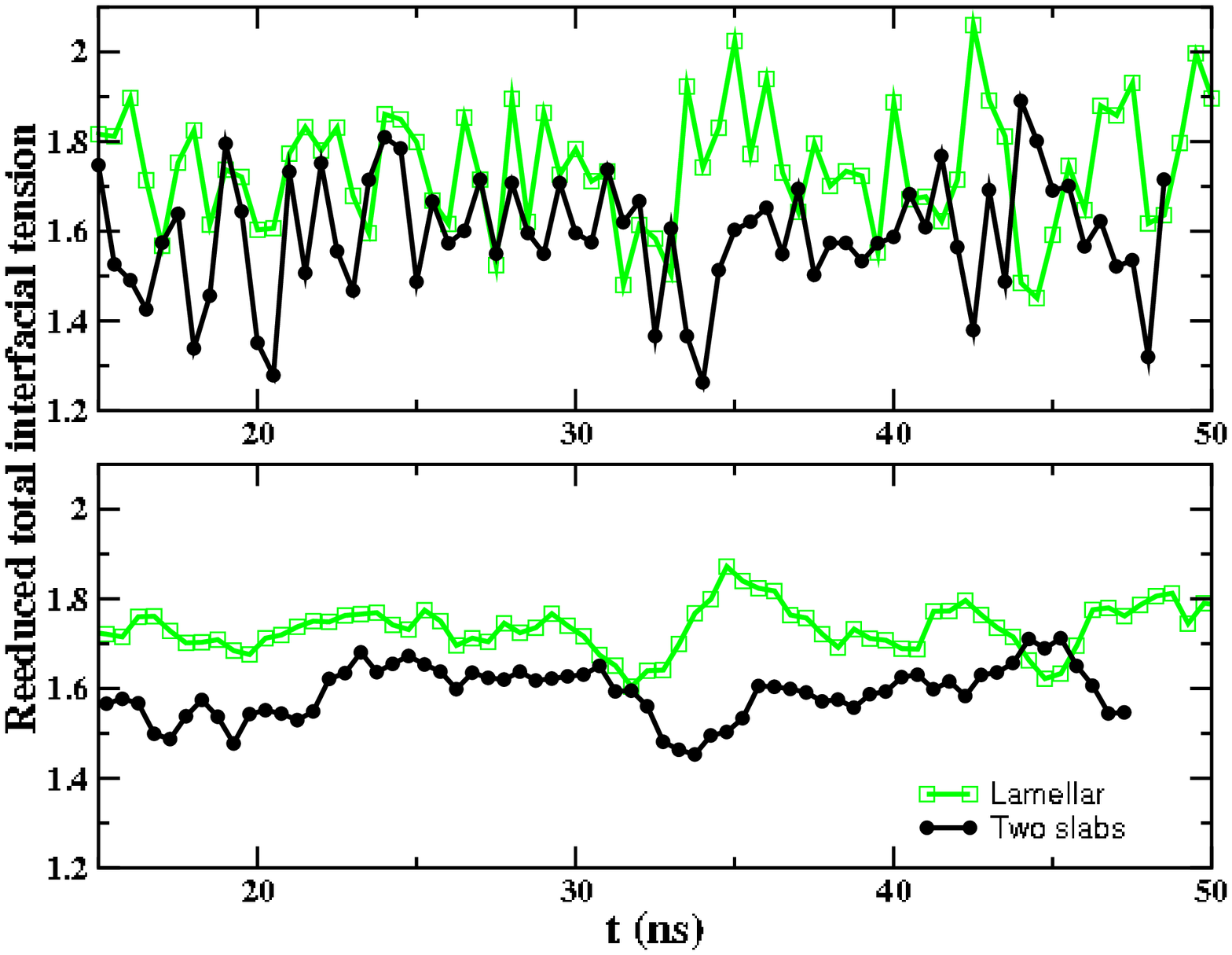}
\end{center}
\caption{Upper figure shows the time evolution of the interfacial energy
and its fluctuations for both, the alternated lamellar and the two
liquid slab state in coexistence with the vapor phase. The large size
of the fluctuations is mainly  attributed to the relatively
high concentration of fluid C at the LL interfaces. In the lower figure
we show the same quantity after a smoothing process has taken place.
As a result one sees that on the average, the lamellar state has a
consistently but slightly higher interfacial tension.}
\label{interfacial}
\end{figure}

Let us now consider a ternary mixture of fluids A, B and C such that
fluids A and B are partially miscible, unlike interactions, 
while the other possible mixtures are miscible, that is, the A-C and B-C
interactions are alike. Therefore, all the matrix elements $\alpha_{i,j}$ of
the interparticle potential, Eq.~\ref{pot}, are equal to unity except those 
corresponding to the A-B interactions for which $\alpha_{A,B}=0.75$,
as defined in section~\ref{models}.
Here the concentrations of fluids A,B and C are approximately one third
of the total concentration,that is, $C_{A}=C_{B}=C_{C}\simeq 33.33\%$.
In order to obtain a well defined lamellar state --at least four lamellae-- 
with these concentrations  we needed a total of 8192 particles of which
2730 particles correspond to fluid C and 2731 for the other two fluids. 
As a consequence of the increase
in the number of particles  the calculations become very demanding and
the the CPU time increases substantially. 
As indicated at the beginning of this section to study the time 
evolution of the mixture we considered an \textit{``ordered"}  and
a \textit{``disordered"}  initial configurations, as described in
what follows.  The \textit{ordered initial configuration} was obtained 
from the two LL slabs in coexistence with the vapor phase of the binary 
mixture at $ T^{*}=0.65 $ and the particles of the third specie were
taken from the two liquid slabs. The \textit{disordered initial 
configuration} is defined as one in which the particles of the three
fluids are randomly placed on the sites of an FCC crystal.
For each initial configuration the mixture was simulated for as 
long as 55~ns. 
Fig.~\ref{laminas3spec} shows the time evolution of the density 
profiles when the simulation initiated from a {\it random configuration}. 
Each panel represents the average of the density profiles over 11~ns. 
One sees that a pretty well defined alternated liquid A-liquid B lamellar 
state with the number of lamellae constant is reached. In addition, one 
observes that the distribution of fluid C is uniform at the liquid phases. 
However, its concentration increases at the LL interfaces. This  
is consistent with a recent study as reported in reference \cite{diaz99}. 
Because of this the interfacial tension of the LL interface decreases
significantly as compared to the corresponding values of the binary mixture,
as will be shown below.
On the other hand, Fig.~\ref{slab3spec} shows the time evolution of
the mixture when the simulation started from an {\it ordered configuration}. 
One sees that the basic structure of the  initial configuration remains stable 
in time with the surfactant uniformly distributed in the liquid phases and an
increase in concentration at the LL interface. This distribution of fluid C is 
similar to that found in the alternated lamellar state. 
The oscillations that are seen in the density profiles are mainly due 
to local fluctuations of the concentration since a rearrangement of 
the fluid particles is taking place. One would expect that after a 
very long time the system will eventually end up  forming two 
separated liquid slabs of the
same width in coexistence with the vapor phase.
At this point we are not able to indicate clearly which one of these 
states corresponds to the equilibrium state because it is very difficult 
to evaluate explicitly the free energy of the system. 
To try to circumvent this difficulty we proceed as in the binary mixture case,
that is, we follow  the time evolution  of the potential energy for both 
structures at $ T^{*}=0.65 $ and the results are plotted in 
Fig.~\ref{upot3spec}. In the upper panel we observe that 
the lines representing the potential
energies are pretty close to each other within the error bars. To see
which system has more potential energy we removed the error bars from
these curves. The curves are plotted in the lower panel of the 
same figure where we observe that for most of the 
time of the simulations the potential energies are essentially the same
within the accuracy of the calculations. This may be an indication 
that both states are approaching the same thermodynamic state. To
try to clarify this issue we proceed further and evaluate the 
interfacial energy per unit area of both states. 
To evaluate the reduced interfacial tension 
we used the Kirwood-Buff formula~\cite{kirwood-buff} that has 
proved to give reasonable estimates~\cite{diaz99}.
The results of this calculation for the liquid-vapor plus liquid-liquid 
interfaces are plotted as a function of time for both configurations in 
Fig.~\ref{interfacial}. In the  upper panel one can observe that the 
interfacial tension shows strong fluctuations in both cases.
To understand the time evolution of $\gamma^{*}$ 
we carried out a smoothing procedure of the data and the results 
are shown in the lower panel of the figure. In doing so we averaged
$\gamma^{*}$ every four points. We observe that for both
configurations the trend in $\gamma^{*}$ is about the same, however, 
the interfacial tension of the lamellar state is systematically above the
the interfacial tension of the two liquid slabs system. The time average 
difference of this quantity is about $0.13$. It is important to point out
that the difference between the total interfacial tensions of these two
states comes mainly from the LL interfaces since the LV interfacial 
tensions are similar and thus cancell out.
The  decrease in $\gamma^{*}$ relative to the binary system is due to the 
surfactant character of the fluid C which concentration is slightly higher at 
the LL interfaces. These results for the potential and interfacial energies 
indicates that in spite of the surfactant character of fluid C, 
the alternated LL lamellar state remains stable for quite a long time.
This happens because there should be an energy barrier that is large 
enough to be overcome by the system when the initial configuration of 
fluids A,B and C is fully random. Therefore, our results strongly 
suggest that the alternated LL lamellar state is a long lived metastable 
state even in the case when fluid C plays the role of a surfactant. 
Of course as one would expect, if the concentration of fluid C 
increased one should reach a concentration at which the lamellar 
state would become into two liquid A liquid B slabs. 
Increasing further the concentration of C  would finally lead to 
a mixed state.
\section{Conclusions.}
\label{conclusions}
We have shown from extensive MD simulations that by properly choosing the
interaction  parameters of the interparticle potentials and the initial 
conditions, long lived metastable alternated  LL lamellar states can be 
achieved in partially miscible binary and ternary mixtures  of LJ fluids. 
These lamellar  states are found to be stable and sharper at relatively 
low temperatures, in the temperature range of the fluid phase 
($78-102~^{\rm o}$K). At temperatures well above $102~^{\rm o}$K we obtain 
a mixed liquid phase  in coexistence with the vapor and the mixture  
reaches relatively easy  the equilibrium state. However, at temperatures 
below this value the liquid phase of the system is demixed and relaxes 
slowly to the equilibrium state. We have also found that for 
\textit{ disordered initial conditions } we obtain long lived LL 
alternated lamellar states. One important question to be answered
in a future research is how is related the number of lamellae with
the number of local equilibria. 
For the ternary mixture that is essentially formed by the partially
miscible binary mixture and a surfactant fluid, we have found that even in
this case pretty stable long lived alternated liquid-liquid lamellar 
structures can be achieved by considering a \textit{disordered the initial 
state} as well. The alternated LL lamellar structures found here are more
or less similar to those obtained in a recent density functional study of 
the LV coexistence of  amphiphile dumbells in coexistence with a LJ 
fluid~\cite{nap00}.  Nonetheless, in this latter case the alternated lamellar 
state constitutes an equilibrium state. Bearing in mind this model one may 
consider the ternary mixture studied in the present paper as one in which 
particles B-C (or A-C) form \textit{effective amphiphile-like molecules} 
interacting with fluid A (or B).  
It is important to stress the fact that here we have 
obtained these LL lamellar structures in coexistence with the vapor phase 
by considering spherical single particle LJ fluids. Nevertheless, these 
appear to be long lived metastable states that are stable and sharper 
in time at relatively low temperatures.
\section{Acknowledgements}
EDH would like to
acknowledge enlightening discussions with F. Forstmann.
This work has been supported by CONACYT research grants Nos. 
L0080-E (EDH), and  25298-E, G32723-E (GRS). 

\newpage
\vspace{2.5in}
\newpage
\hbox{     }
\hbox{         }
\vspace{1.5in}
\newpage
\vspace{1.5in}
\newpage
\vspace{1.5in}
\newpage
\vspace{1.5in}
\begin{figure}
\begin{center}
\includegraphics[width=3.25in,height=3.0in]
{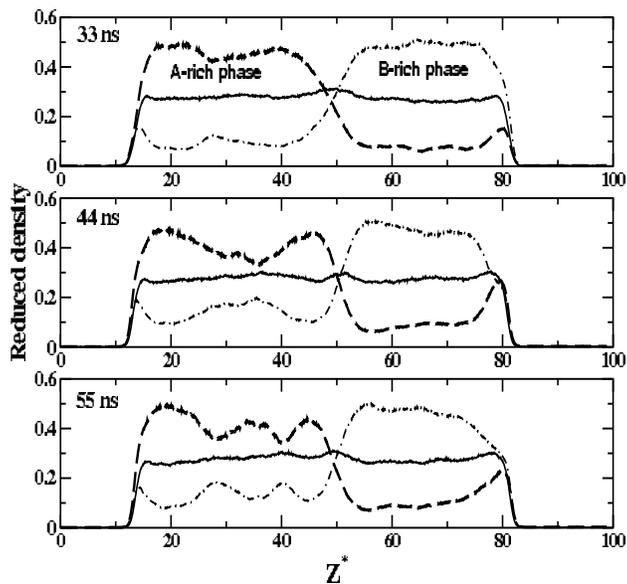}
\end{center}
\caption{Time evolution of the density profiles of the lamellar state at
$T^{*}= 0.65$. Averages are calculated using three subsequent blocks of one
million time steps each. Here the system started from an ordered
configuration. Notice that in the third block the density profiles
of fluid A show oscillations.}
\label{slab3spec}
\end{figure}
\newpage
\vspace{1.5in}

\begin{thebibliography}{3}
\bibitem{dagamma1}
M.M. Telo da Gamma and R. Evans, Mol. Phys. {\bf 41}, 1091 (1980),
{\it ibid}, Faraday Symp. Chem. Soc. {\bf 16}, 45 (1981).
\bibitem{tarazona1}
P. Tarazona, M.M. Telo da Gamma and R. Evans, Mol. Phys. {\bf 49},
283 (1983), {\it ibid}, {\bf 49}, 301 (1983).
\bibitem{jlee1}
J. Lee, M.M. Telo da Gamma and K. E. Gubbins, Mol. Phys. {\bf 53},
1113 (1984), {\it ibid} J. Phys. Chem. {\bf 89}, 1514 (1985).
\bibitem{almeida}
B. S. Almeida and M.M. Telo da Gamma J. Phys. Chem. {\bf 93},4132
(1989).
\bibitem{wendall}
M. Wendland, Fluid Phase Equilibria {\bf 141}, 25 (1997),
{\it ibid}, {\bf 147}, 105 (1998).
\bibitem{fortsman97}
Stanislav Iatsevitch and Frank Forstmann, J. Chem. Phys. 107, 6925, (1997).
\bibitem{nader2002}
N. Lotfohalli, Mohammad, Modarres, Hamid, J. Chem. Phys. {\bf 166},
2487 (2002).
\bibitem{chapela}
G. A. Chapela, G. Saville, S. M. Thompson, and J. S. Rowlinson,
J. Chem. Soc., Faraday Trans. 2 {\bf 73}, 1133 (1977).
\bibitem{salomons}
E. Salomons and M. Mareschal, J. Phys.: Condens. Matter {\bf 3}, 3645
(1991), {\it ibid}, {\bf 3}, 9215 (1991).
\bibitem{holcomb}
C. D. Holcomb, P. Clancy, S. M. Thompson, and J. A. Zolweg,
Fluid Phase Equilibria {\bf 75}, 185 (1992); {\it ibid}
{\bf 88}, 303 (1993); C. D. Holcomb, P. Clancy, and J. A. Zolweg,
Mol. Phys. {\bf 78}, 437 (1993).
\bibitem{li95}
Li-Jen Chen, J. Chem. Phys. {\bf 103}, 10214 (1995).
\bibitem{mecke99}
M. Mecke and J. Winkelmann, J. Chem. Phys. {\bf 110}, 1188 (1999).
\bibitem{alejandre99}
Andrij Trokhymchuck and Jos\'e Alejandre, J. Chem. Phys. {\bf 111}, 8510, (1999).
\bibitem{diaz99}
Enrique D\'\i az-Herrera, Jos\'e Alejandre, Guillermo Ram\'\i rez-Santiago and
F. Forstmann,  J. Chem. Phys. {\bf 110},
8084, (1999).
\bibitem{strey95}
R. Strey, Y. Viisanen, and P. E. Wagner, J. of Chem. Phys.
{\bf 103}, 4333 (1995).
\bibitem{guerra98}
C. Guerra, A. M. Somoza and M.M. Telo da Gamma, J. Phys. Chem. {\bf 109},
1152 (1998).
\bibitem{endo2000}
H. Endo, M. Mihailescu, M. Monkenbusch, J. Allgaier, 
G. Gompper, D. Richter, B. Jakobs, T. Sottmann, 
R. Strey and I. Grillo, J. Chem. Phys. {\bf 115}, 580 (2001).
\bibitem{isamu2001}
I. Kusaka, D. W. Oxtoby, J. Chem. Phys. {\bf 115}, 4883 (2001).
\bibitem{nijmeijer88}
M.J.P. Nijmeijer, A. F. Bakker, C. Bruin, and J. H. Sikkenk,
J. Chem. Phys. {\bf 89},3789 (1988).
\bibitem{toxvaerd95}
S. Toxvaerd and J. Stecki, J. Chem. Phys. {\bf 102}, 7163 (1995).
\bibitem{stecki95}
J. Stecki and S. Toxvaerd,  J. Chem. Phys. {\bf 103}, 4352 (1995);
{\it ibid.} J. Chem. Phys. {\bf 103}, 9763 (1995).
\bibitem{padilla96}
P. Padilla and J. Stecki, J. Chem. Phys. {\bf 104}, 7249 (1996).
\bibitem{wilding97}
N. B. Wilding, Phys. Rev. E {\bf 55}, 6624 (1997).
\bibitem{wilding98}
N. B. Wilding, F. Schmid, P. Nielaba, Phys. Rev. E {\bf 58}, 2201 (1998).
\bibitem{gang00}
Jian-Gang Weng, Seungho Park, Jennifer R. Lukes, and Chang-Lin Tien, 
J. Chem. Phys. {\bf 113}, 5917, (2000).
\bibitem{kirwood-buff}
J. G. Kirwood and F. P. Buff, J. Chem. Phys. {\bf 17}, 338 (1949).
\bibitem{nap00}
Ismo Napari and Ari Laaksonen, Phys. Rev. Lett. {\bf 84}, 2184 (2000).
\end{thebibliography}
\end{document}